# Laser Polarization Effects in Supercontinuum Generation

Alok Srivastava[a] and Debabrata Goswami[a,b]

[a] Tata Institute of Fundamental Research, Mumbai – 400005

[b] Indian Institute of Technology, Kanpur – 208016

We present experimental evidence of a fourth order process in electric field in supercontinuum generation. We also show laser induced polarization preference in the supercontinuum generating media. These results have become possible through the choice of isotropic and anisotropic samples interacting with ultrashort laser pulses of changing ellipticity. Laser polarization emerges as an important control parameter for the highly nonlinear phenomenon of supercontinuum generation.

42.65.Ky, 42.65.Re

Supercontinuum (SC) generation is a highly nonlinear process which occurs when an intense ultrashort laser pulse emerges through a condensed or a gaseous medium with a much wider coherent spectral content[1]. SC generation has been explored for many years both due to its fascinating physics and applications involving various light-matter interactions, spectroscopy, etc.[2,3] However, there are situations where SC generation must be suppressed and our studies show that it can be achieved by the changing the polarization of the incident laser pulses[4,5]. We establish that circularly polarized laser pulses produce smaller SC as compared to linearly polarized pulses irrespective of the nature of the sample. The intensity dependence of SC generation shows that the suppression rate of SC depends on the intensity of the input pulses.

The work presented in this paper focuses on the effect of the sample anisotropy on the SC and this, to our understanding, is the first experimental study which shows the presence of higher order optical processes in the SC generation. Our experimental results also show evidence of laser induced polarization preference in supercontinuum generating media.

For the SC generation experiments, we use linearly polarized laser pulses from a chirped pulse multipass amplified Ti:Sapphire laser (Odin, Quantronix Inc.) which produces femtosecond pulses of 50fs, 800nm, 1mJ/pulse at 1 kHz repetition rate. The Odin amplifier is operated under the best stability conditions (Power instability < 2%), which is monitored continuously throughout the experiment. A small fraction of main beam from the amplifier is passed through a pair of spatial filters (iris diaphragms) to produce a uniform circular beam profile which, we found, is necessary for a stable SC. This spatially filtered beam, having power of about 40mJ, is focused with a 50cm lens into the sample placed about 6cm before the focus. In this arrangement the spot-size of the beam is ≈0.6mm at the sample giving an intensity of the order $10^{10}$W/cm$^2$ and stable SC from the liquids in the cuvette or the solid samples. A quarter-wave plate is used to vary the incident laser polarization from linear through elliptical to circular and vice-versa. Though the laser peak powers used in these experiments are above the self-focusing thresholds for the samples used, the large spot-size at the sample surface ensures that SC does not occur until the self-focusing process brings down the spot-size inside the sample to the threshold intensities. We also ensured that there is no SC from the cuvette alone or in air at the focal point of the laser without a cuvette or the solid sample. We have chosen optically transparent samples to avoid self-absorption of SC by the generation medium itself. The SC is collected with lenses into a monochromator (Acton Research Inc. SP150) for the spectral contents while for the integrated measurements we have used a PMT (Hamamatsu 1P28). The polarization dependence of SC was obtained from the integrated measurements with the help of a PMT. The integrated SC is measured as a function of the incident laser polarization which is varied with the help of a first-order $\lambda/4$ waveplate. For these measurements we cut off the fundamental and the IR light completely with a pair of 720nm cut-off short-pass filters (Melles Griot 03SWP614). This ensures that only the UV and visible part of the supercontinuum light is recorded. The output polarization of SC, measured using a Glan-Laser polarizer with an effective range of 400-1000nm, is found to be the same as the input laser polarization for both the linear and circular cases. We do not make any polarization related measurements through the monochromator as the grating response depends upon the polarization while PMT has no such dependence. We also measure a linear component of the incident laser beam through a Glan-Laser polarizer for various positions of the $\lambda/4$ plate without any sample. The effect of sample anisotropy on SC is studied by the rotation of the plane of incident laser polarization using a $\lambda/2$ plate for anisotropic samples. The data points corresponding to each polarization setting are averaged over 1000 shots. All the experiments have been repeated under identical conditions several times resulting in reproducible data. This also ensured that there was no sample damage during these experiments. This data-acquisition procedure gives error-bars smaller than the symbol-size used in Figs. 1 and 2.

The overall spectral characteristics of SC from different isotropic samples of water, acetone, methanol, fused silica etc. are found to be similar and independent of the polarization of incident laser pulses. However, the generated SC shows a strong dependence on the ellipticity of the incident laser pulse for all the samples, isotropic or otherwise. Typical ellipticity dependence of SC for a few isotropic samples, such as, acetone, water and 2-S-Butanol (chiral liquid) is shown in Fig.1. We also plot SC generated from the anisotropic sample of sapphire in the same figure. We find from Fig.1 that irrespective of the nature of these samples, the generated SC show similar ellipticity dependence.

We also measured a linear component of the elliptically polarized laser pulse through a polarizer (analyzer) as a function of ellipticity of the incident beam. For this measurement, the 800nm filters, spatial filters and the focusing lenses are removed and the polarizer is set for the maximum transmission of linearly polarized incident beam. The ellipticity of the incident beam is changed with the help of a $\lambda/4$ plate placed before the analyzer. The linear component as well the SC measurements have been done with the same $\lambda/4$ plate rotations and therefore we plot it with all the SC data in Fig.1. We observe a complete correspondence between the two distinct data sets which is very surprising. We find that a cosine squared function of the angle of rotation ($\theta$) of quarter-waveplate: $A + B\cos^2(2\theta + \pi/2)$ (solid fit curve in Fig.1) is able to fit the ellipticity dependence of the linear component and all of the SC data.

A closer inspection into the anisotropic sample (sapphire) data shows a small deviation from the above cosine squared function. This additional effect is an order of magnitude lower than the ellipticity effect and is barely evident in Fig.1. However, we find that the anisotropic effect on SC can be explored independently when the incident plane of polarization of the input laser pulses are rotated with the help of a λ/2 plate as shown in Fig.2. These results can be fitted (solid line in Fig.2) by: $A + B\cos^4(\phi + \pi/2)$, where φ is the rotation of the plane of polarization of the incident laser pulses.

The optical response of system is: $P(t) = \chi^{(1)}E(t) + \chi^{(2)}E^2(t) + \chi^{(3)}E^3(t) + \ldots$. For the isotropic samples, even order effects are not present due to symmetry arguments. Thus, the $\chi^{(3)}$ term is responsible for SC in isotropic sample. For the anisotropic samples, however, both the even and odd functions are present. If the $\chi^{(2)}$ term is more important than the $\chi^{(3)}$ term, then the effect of the $\chi^{(2)}$ term would be predominant as compared to that of the $\chi^{(3)}$ term, as is the case in the second harmonic generation (SHG) in barium beta-borate (BBO) crystal. In order to confirm this effect of anisotropy, we also measured SC on a 1mm thick BBO crystal, cut for doubling 800nm, under the non-phase matched condition and found it to be the same as in Fig.2. Since the material coefficients are the same for the nonlinear terms between SHG and SC generation, it is easy to argue that the much smaller effect of polarization rotation as compared to that of changing ellipticity cannot be attributed to the $\chi^{(2)}$ term. From this argument we conclude that the observed effect of anisotropy in SC is actually due to the $\chi^{(4)}$ term.

The sample invariance of the ellipticity dependence of SC and the observed polarization preference which can be fitted to a single cosine squared function are very intriguing. Our results on the anisotropic samples provide the first direct experimental evidence of the presence of higher than $\chi^{(3)}$ process in SC. We also provide a method to control of both third and fourth order processes independently in the same experiment with the help of two different control parameters. The ellipticity dependence of SC in isotropic samples offers a simple procedure for its control and we have experimentally shown that SC generation is suppressed with circularly polarized light.

The laser system used in this research is funded by the Ministry of Communication, Information and Technology, Government of India.

**Figure Captions**

**Fig.1. Plot of SC generation for different samples as a function of the angle of rotation of quarter-waveplate ($\theta$). Ellipticity dependence of the incident laser beam through an analyzer also plotted. All data fits to a cosine squared function (solid line).**

**Fig.2. Anisotropy effects in SC is demonstrated in sapphire when the generated SC is plotted as a function of rotation ($\phi$) of the plane of polarization of the incident laser pulse. A cosine to the power four function fits most of the data (solid line).**

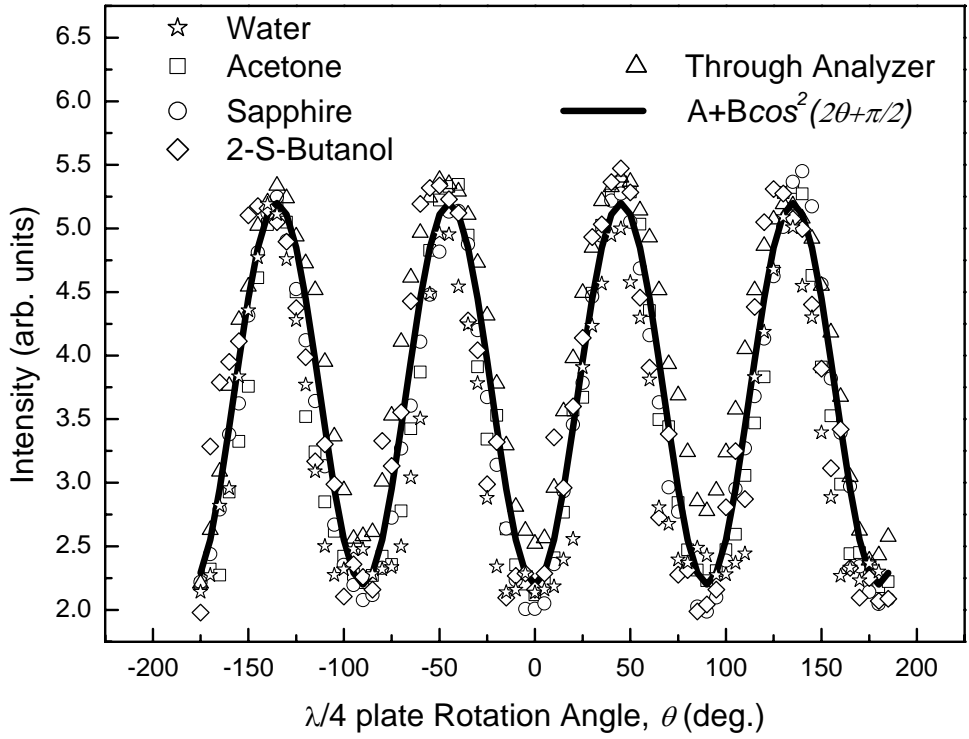

Fig.1

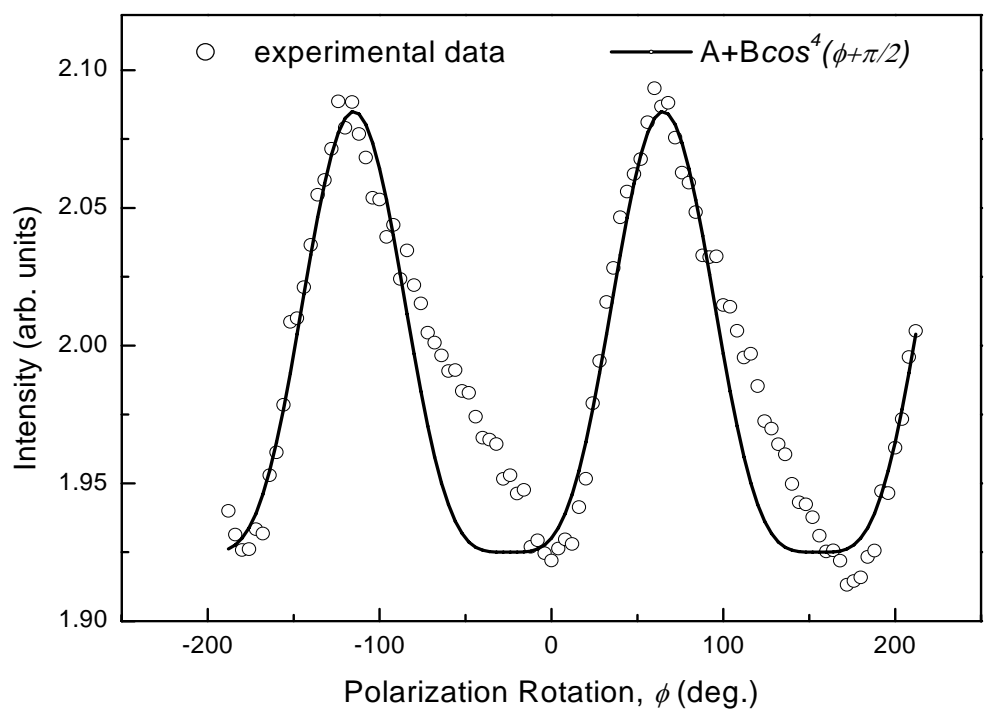

Fig.2